\documentclass[preprint,12pt,review]{elsarticle}

\usepackage{amssymb}
\usepackage{amsmath}

\journal{arXiv.org}

\begin{document}

\begin{frontmatter}



\title{Single vector leptoquark production at hadron colliders due to direct lepton--gluon interaction}


\author{I. Alikhanov\corref{cor1}}

\cortext[cor1]{{\it Email address: {\tt ialspbu@gmail.com}}}

\address{Institute for Nuclear Research of the Russian Academy of Sciences,
60-th October Anniversary pr. 7a, Moscow 117312, Russia}

\begin{abstract}
The cross section of single vector leptoquark production in direct lepton--gluon interaction is calculated. 
In a model independent analysis an effective Lagrangian describing the  most general $C$ and $P$ conserving coupling of the leptoquarks to gluons is considered. An analytical expression is derived for the cross section in  the case of anomalous vector leptoquark
couplings to the gluon field. The cross sections of inclusive production of~($eq$)-, ($\mu q$)- and ($\tau q$)-type vector leptoquarks in $ep$ and $pp$ collisions due to direct lepton--gluon interaction are evaluated for~$q=u, b, t$. Dependences of the cross sections on the anomalous couplings are investigated. The obtained results can be useful for studies at $ep$ colliders and the LHC. 
\end{abstract}


\end{frontmatter}

\section{Introduction}
\label{intro}

Numerous theories beyond the Standard Model lead to the possibility of quark--lepton interactions mediated by bosons called leptoquarks~\cite{lep1,lep2,lep3,lep4,lep5,lep6,lep7}. 

The analysis of this Letter is based on the Buchm\"uller--R\"uckl--Wyler (BRW) leptoquark model~\cite{leptoquark} within which the baryon and lepton numbers are conserved and the corresponding Lagrangian is symmetric under $SU(3)_C\times SU(2)_L\times U(1)_Y$ transformations so that the leptoquarks carry the electric and weak charges and also appear to be color triplets. The latter means that the leptoquarks can couple to the gauge bosons of the Standard Model~\cite{blumlein1993,montalvo,vertex2}. The bosonic couplings of scalar leptoquarks are determined completely, while for the case of vector leptoquark states there is an ambiguity in determination of the Yang--Mills type couplings depending on the nature of the vector leptoquarks. If the leptoquarks are the gauge bosons of an extended gauge group, being thus fundamental objects, then the couplings are completely fixed by the gauge invariance. However, if they are low energy manifestations of a more fundamental theory at a higher energy scale, then the interaction with the gauge bosons is described by an effective Lagrangian. In this case there may be a number of parameters inducing anomalous interactions. Therefore, both possibilities of interaction of the vector leptoquarks should be investigated. 

Searches for leptoquarks have been carried out at $e^+e^-$~\cite{exp1}, $ep$~\cite{exp2}, $p\bar p$~\cite{exp3} and $pp$ colliders~\cite{exp4,exp5}.

The present Letter studies single vector leptoquark production in direct lepton--gluon interaction and generalizes the analysis of~\cite{spira,mine} by considering the most general $C$ and $P$ conserving coupling of gluons to the leptoquarks. The cross sections of production of ($eq$)-, ($\mu q$)- and ($\tau q$)-type vector leptoquarks due to direct lepton--gluon interaction at the LHC is also investigated~for $q=u,b,t$. 

\section{The coupling of vector leptoquarks to gluons}
\label{model}

Let us consider single vector leptoquark ($V$) production in direct interaction of a left/right polarized lepton with gluons appearing in the BRW model~\cite{spira}:

\begin{equation}
l_{L/R}+\text{g}\rightarrow q+V,\label{reac2}
\end{equation} 

where $l=e,\mu,\tau$.

The leading order Feynman diagrams contributing to this process are presented in Fig.~\ref{fig1} in which one encounters the coupling of vector leptoquarks to gluons mentioned above.  The most general effective Lagrangian describing vector leptoquark--gluon interaction has the form \cite{blumlein_main}:

\begin{eqnarray}
{\cal L}_V^\text{g}
= \sum_{vectors} \left \{ -\frac{1}{2} G^{i \dagger}_{\mu \nu}
G^{\mu \nu}_i + m_V^2 V_{\mu}^{i \dagger} V^{\mu}_i\right.\nonumber\\
\left.-ig_s \left [ (1 - k_G)
V_{\mu}^{i \dagger}
t^a_{ij}
V_{\nu}^j
{\cal G}^{\mu \nu}_a
+ \frac{\lambda_G}{m_V^2} G^{i\dagger}_{\sigma \mu}
t^a_{ij}
G_{\nu}^{j \mu} {\cal G}^{\nu \sigma}_a \right ] \right \}.\label{lagrange}
\end{eqnarray}

Here,
$g_s$ is the  strong coupling
constant, $t^a_{ij}$ are the generators of the group $SU(3)_C$,
$m_V$ is the leptoquark mass, $k_G$ and
$\lambda_G$ are the anomalous couplings.

The field strength tensors of the  gluon and vector
leptoquark fields are
\begin{eqnarray}
{\cal G}_{\mu \nu}^a  &=& \partial_{\mu} {\cal A}_{\nu}^a
 - \partial_{\nu}
{\cal A}_{\mu}^a + g_s f^{abc} {\cal A}_{\mu b} {\cal A}_{\nu c},
 \nonumber\\
G_{\mu \nu}^{i}
 &=& D_{\mu}^{ik}
 V_{\nu k} - D_{\nu}^{ik} V_{\mu k},
\end{eqnarray}
with the covariant derivative given by
\begin{equation}
D_{\mu}^{ij} = \partial_{\mu} \delta^{ij} - i g_s
t_a^{ij}
 {\cal A}^a_{\mu}.
\end{equation}

The parameters $k_G$ and $\lambda_G$ are related
to the anomalous "magnetic" moment
$\mu_V$ and "electric" quadrupole moment $q_V$
of the leptoquarks in the color field and assumed to be real

\begin{eqnarray}
\mu_{V,G} = \frac{g_s}{2 m_{V}} \left ( 2 -
k_G + \lambda_G \right ), \nonumber\\
q_{V,G} = - \frac{g_s}
{m_{V}^2} \left (1 - k_{G} - \lambda_{G} \right ).
\end{eqnarray}

To keep the analysis model independent,
it is also assumed that these quantities are independent.

The Lagrangian (\ref{lagrange}) leads to the following Feynman rule for the trilinear $VV\text{g}$ vertex shown in Fig.~\ref{fig2}:

\begin{eqnarray}
V_{\beta \sigma \alpha}^{VV\text{g}, aij}(p_1, p_2, p_3) &=&
ig_s(t^a)^{ij} \left[ \widehat{V}_{\beta \sigma \alpha} +
k_G \widehat{V}_{\beta \sigma \alpha}^{k}
+ \frac{\lambda_G}{m_V^2} \widehat{V}_{\beta \sigma \alpha}
^{\lambda} \right],\label{vertex_gluon}
\end{eqnarray}
where
\begin{eqnarray}
\widehat{V}_{\beta \sigma \alpha}(p_1, p_2, p_3) &=&
  (p_3 - p_1)_{\beta} g_{\sigma \alpha}
+ (p_2 - p_3)_{\alpha} g_{\beta \sigma}
+ (p_1 - p_2)_{\sigma} g_{\alpha \beta},   \\
\widehat{V}_{\beta \sigma \alpha}^{k} (p_1, p_2, p_3) &=&
p_{1 \beta} g_{\sigma \alpha} - p_{1 \sigma} g_{\beta \alpha},
\\
\widehat{V}^{\lambda}_{\beta \sigma \alpha}(p_1, p_2, p_3) & = &
(p_2\cdot p_3) (p_{1 \beta} g_{\sigma \alpha} -  p_{1 \sigma} g_{\beta
  \alpha})
+ (p_1\cdot p_3) (p_{2 \sigma} g_{\beta \alpha} -  p_{2 \alpha} g_{\beta
\sigma}) \nonumber\\
& + & (p_1\cdot p_2) (p_{3 \alpha} g_{\beta \sigma} -  p_{3 \beta}
g_{\sigma \alpha})
+ p_{2 \alpha} p_{3 \beta} p_{1 \sigma} - p_{2 \sigma}
p_{3 \alpha} p_{1 \beta}.
\end{eqnarray}

The Feynman rules for the rest two vertices, $eqV$ and $qq\text{g}$, needed to calculate the cross section are much simpler and given 
in Fig.~\ref{fig3}.

\section{The cross section for direct lepton--gluon interaction\label{calcs}}

\begin{figure}
\centering
\resizebox{1.0\textwidth}{!}{%
\includegraphics{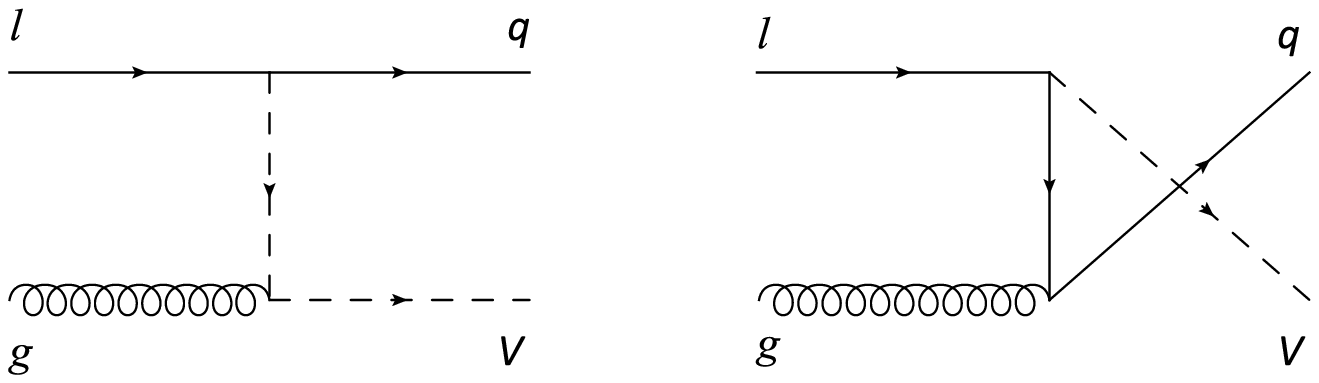}
}\caption{}
\label{fig1}
\end{figure}

\begin{figure}
\centering
\resizebox{0.4\textwidth}{!}{%
\includegraphics{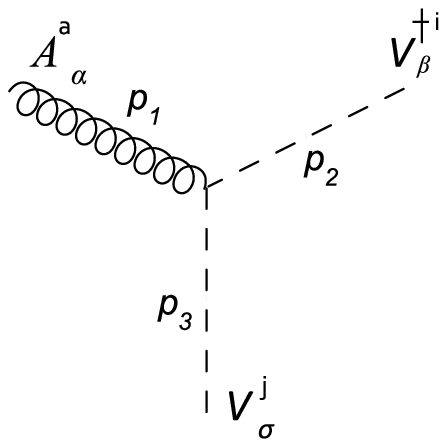}
}\caption{}
\label{fig2}
\end{figure}

\begin{figure}
\centering
\resizebox{0.6\textwidth}{!}{%
\includegraphics{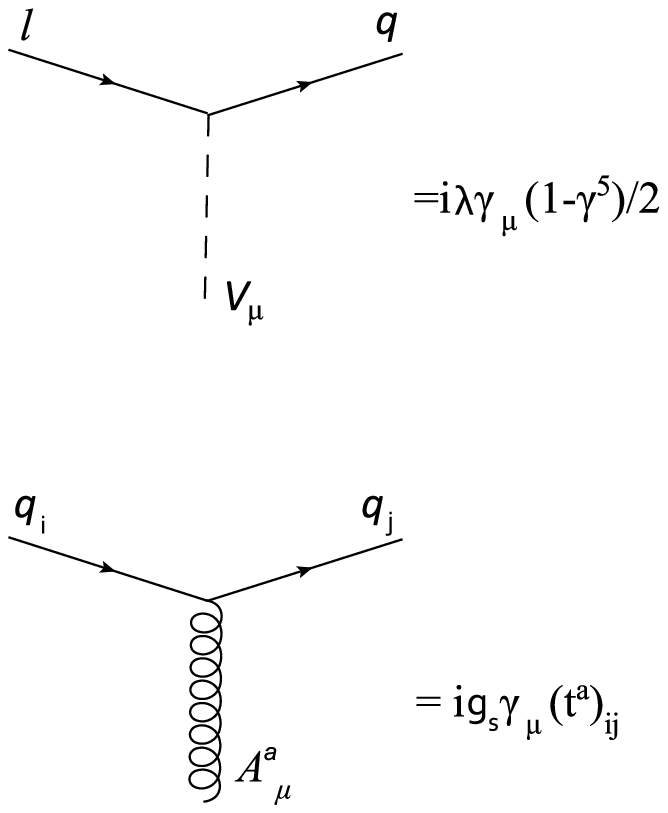}
}\caption{}
\label{fig3}
\end{figure}

The cross section of the reaction (\ref{reac2}) is calculated using the diagrams from Fig.~\ref{fig1} with the corresponding Feynman rules given by~(\ref{vertex_gluon}) and Fig.~\ref{fig3}. The result in the limit of the vanishing lepton mass reads

\begin{eqnarray}
\sigma_{l\text{g}}(s)=\frac{\lambda^2\alpha_s}{8m_V^2s^3}\left\{\frac{p\sqrt{s}}{4m_V^2} \left[\sum_{i=1}^{3}f_i(s,k_G,\lambda_G)+4m_V^2(16m_V^4-5m_q^4)\right]\right.\nonumber\\\left.+\left[\sum_{i=4}^{6}f_i(s,k_G,\lambda_G)-8m_V^4(s-m_V^2)-2m_q^6\right] \log
   \left[\frac{\left(p^2+m_V^2\right)^{1/2}+p}{m_V}\right]\right.\nonumber\\\left.+2\left[f_7(s,k_G,\lambda_G)+2m_V^2(s^2-2m_V^2s+2m_V^4)-m_q^6\right] 
   \log   \left[\frac{\left(p^2+m_q^2\right)^{1/2}+p}{m_q}\right]\right\},\nonumber\\\label{cross_main}
\end{eqnarray}

where $\lambda$ is the coupling constant corresponding to the $lqV$ vertex, $\alpha_s=~g^2_s/4\pi$,

\begin{equation}
f_1(s,k_G,\lambda_G)=m_q^2(m_V^2s(k_G^2-2(k_G+12)\lambda_G+28k_G-44)-3\lambda_G^2s^2-44m_V^4),
\end{equation}

\begin{equation}
f_2(s,k_G,\lambda_G)=m_V^2s^2(k_G^2-2k_G(9\lambda_G+16)+13\lambda_G^2+32),
\end{equation}

\begin{equation}
f_3(s,k_G,\lambda_G)=-m_V^4s(k_G-\lambda_G)^2+\lambda_G^2s(m_q^4+4s^2),
\end{equation}

\begin{equation}
f_4(s,k_G,\lambda_G)=m_q^2(s^2(\lambda_G^2-k_G^2-3k_G+1)+m_V^2s(k_G+2\lambda_G+4)+6m_V^4),
\end{equation}

\begin{equation}
f_5(s,k_G,\lambda_G)=s^3(k_G^2+2k_G\lambda_G-3\lambda_G^2)-m_V^2s^2(k_G^2-2k_G(\lambda_G+2)+\lambda_G^2),
\end{equation}

\begin{equation}
f_6(s,k_G,\lambda_G)=m_q^4s(3k_G-2\lambda_G+1)-12m_q^4m_V^2,
\end{equation}

\begin{equation}
f_7(s,k_G,\lambda_G)=2m_q^4((k_G+1)s-3m_V^2)-m_q^2((k_G-1)s^2-3m_V^4+m_V^2s)
\end{equation}

and $p=\sqrt{\left(s-(m_V+m_q)^2\right)\left(s-(m_V-m_q)^2\right)}/2\sqrt{s}$ is the center-of-mass momentum of any of the final state particles. 

One can verify that (\ref{cross_main}) is in a very good numerical agreement with the corresponding calculations of~\cite{spira} at $k_G=1$, $\lambda_G=0$ (the minimal vector coupling) and exactly reducible to the result from~\cite{mine} by setting $k_G=\lambda_G=0$ (Yang--Mills type coupling) as well. 

Note that from (\ref{cross_main}) one can also obtain the exact expression for the leading order cross section of $W$ boson production in neutrino--photon scattering $\nu_l+\gamma\rightarrow l+W$~\cite{seckel} by performing the following substitutions~\cite{mine}:

\begin{equation}
\lambda\rightarrow\frac{g}{\sqrt{2}},\: \frac{\alpha_s}{2}\rightarrow\alpha,\:m_V\rightarrow m_W,\:m_q\rightarrow m_l, k_G=\lambda_G=0, \label{replace1}
\end{equation}

where $g$ is the coupling of the weak charged current (related to the Fermi coupling constant $G_F$ by $G_F=\sqrt{2}g^2/8m_W^2$, $\alpha$ is the fine structure constant. The coefficient of $\alpha_s$ is the color factor equal to $1/2$  for the diagrams in Fig.~\ref{fig1}.

\section{Single vector leptoquark production in lepton--proton collisions}

A standard convolution of the cross section (\ref{cross_main}) with the gluon distribution in the proton gives the cross section of inclusive vector leptoquark production measurable in lepton--proton collisions:  

\begin{equation}
\sigma_{lp}(s)=\int_{x_0}^1dx\,\text{g}(x,\hat s)\sigma_{l\text{g}}(xs),\label{convolution}
\end{equation}

where $\text{g}(x,\hat s)$ is the gluon distribution function, $\sqrt{\hat s}$ is the total energy of the subprocess,
$x_0=(m_V+m_q)^2/s$. 

\begin{figure}
\centering
\resizebox{0.9\textwidth}{!}{%
\includegraphics{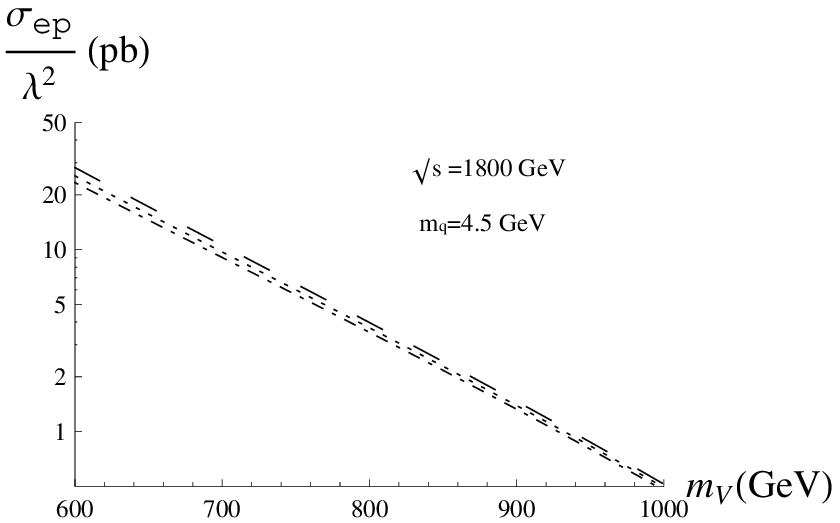}
}\caption{}
\label{fig4}
\end{figure}

\begin{figure}
\centering
\resizebox{1.2\textwidth}{!}{%
\includegraphics{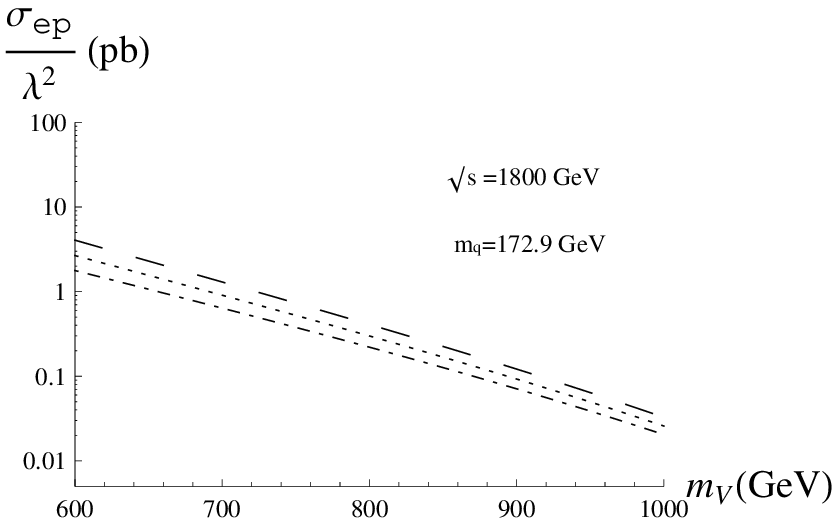}
}\caption{}
\label{fig5}
\end{figure}

Figures~\ref{fig4} and~\ref{fig5} show the cross sections for the production of the ($eb$) and ($et$)-type vector leptoquarks in $ep$ collisions evaluated using (\ref{convolution}) with the gluon distribution function adopted from CTEQ5~\cite{gluon_distr} at a few fixed values of the anomalous couplings.

\begin{figure}
\centering
\resizebox{0.9\textwidth}{!}{%
\includegraphics{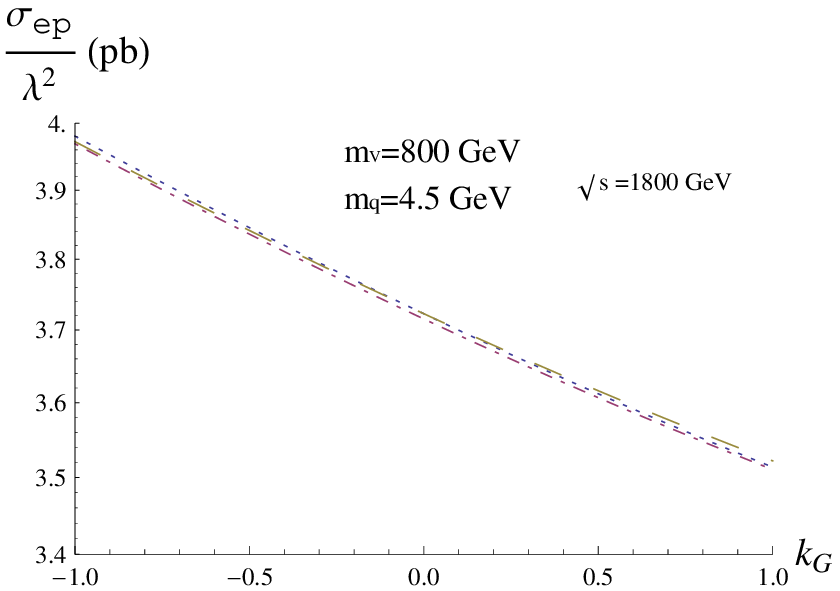}
}\caption{}
\label{fig6}
\end{figure}

\begin{figure}
\centering
\resizebox{0.9\textwidth}{!}{%
\includegraphics{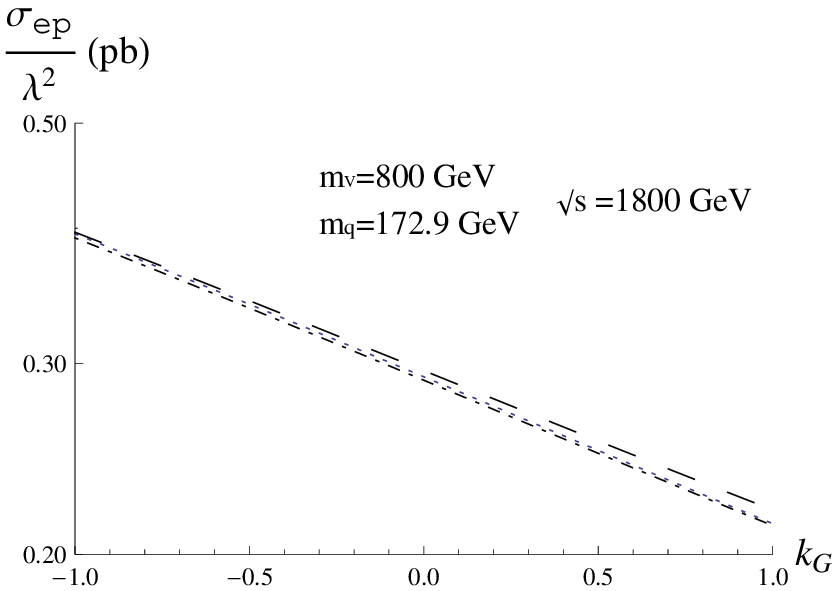}
}\caption{}
\label{fig7}
\end{figure}

\begin{figure}
\centering
\resizebox{0.9\textwidth}{!}{%
\includegraphics{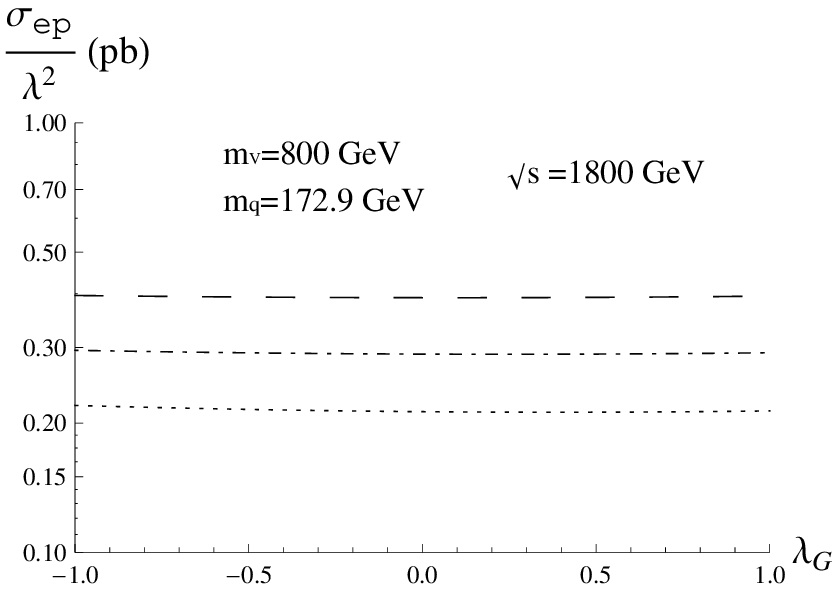}
}\caption{}
\label{fig8}
\end{figure}

Dependences of the cross section (\ref{convolution}) on the anomalous couplings are also investigated in the ranges $|k_G|$, $|\lambda_G|\leq1$ covering both the particular cases of the minimal vector and the Yang--Mills couplings. The corresponding results for production of ($eb$) and ($et$)-type vector leptoquarks of mass 800~GeV at $\sqrt{s}=1800$~GeV are presented in Figs. \ref{fig6},~\ref{fig7} and~\ref{fig8}. 

\section{Single vector leptoquark production in proton--proton collisions}

\begin{figure}
\centering
\resizebox{0.8\textwidth}{!}{%
\includegraphics{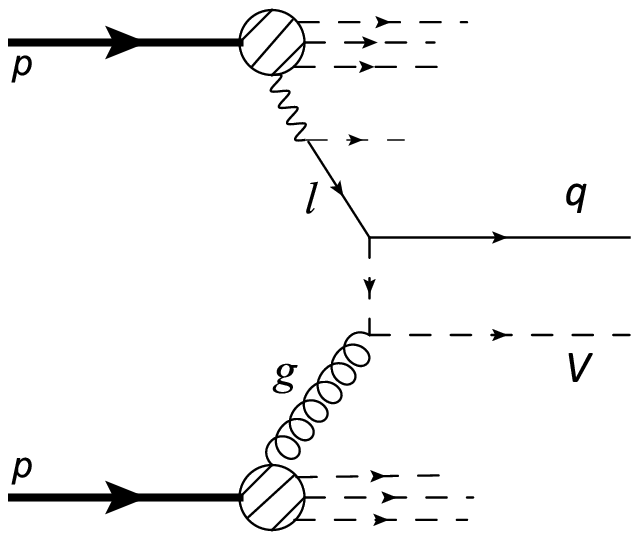}
}\caption{}
\label{fig9}
\end{figure}

Leptoquarks can be produced singly at $pp$ colliders such as the LHC by splitting photons emitted from the proton
beam into lepton pairs~\cite{zerwas94}. One of the leptons can then collide with a gluon from the other proton beam and thereby produce a leptoquark. This mechanism is schematically illustrated in Fig~\ref{fig9}. The corresponding  cross section is then given by

\begin{equation}
\sigma_{pp}(s)=\int_{x_0}^1dx_1\int_{x_0/x_1}^1dx_2\,f_{l/p}(x_1,\hat s)\text{g}(x_2,\hat s)\sigma_{l\text{g}}(x_1x_2s).\label{convolution2}
\end{equation}

Here $f_{l/p}(x_1,\hat s)$ is the equivalent lepton spectrum of the proton which can be found as 

\begin{equation}
f_{l/p}(x,\hat s)=\int_{x}^1\frac{dz}{z}f_{l/\gamma}(x/z,\hat s)f_{\gamma/p}(z,\hat s),\label{convolution3}
\end{equation}

where $f_{\gamma/p}(x,\hat s)$ is the equivalent photon spectrum of the proton and $f_{l/\gamma}(x,\hat s)$ is the photon splitting rate to lepton pairs of mass $m_l$~\cite{split1,split2,split3,split4}

\begin{equation}
f_{l/\gamma}(x,\hat s)=\frac{\alpha}{2\pi}\left[x^2+(1-x)^2\right]\log{\frac{\hat s}{m^2_l}}.
\end{equation}

In the present analysis, for $f_{\gamma/p}(x,\hat s)$, the inelastic photon content of the proton from~\cite{gluk95} is adopted. It is also set $\hat s=m^2_V$ in all the used partonic distribution functions~\cite{zerwas94}. 

\begin{figure}
\centering
\resizebox{0.9\textwidth}{!}{%
\includegraphics{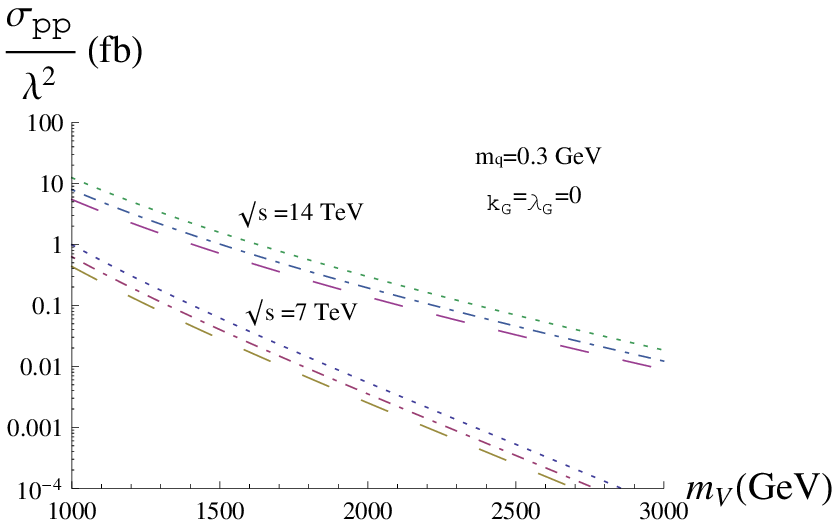}
}\caption{}
\label{fig10}
\end{figure}

\begin{figure}
\centering
\resizebox{0.9\textwidth}{!}{%
\includegraphics{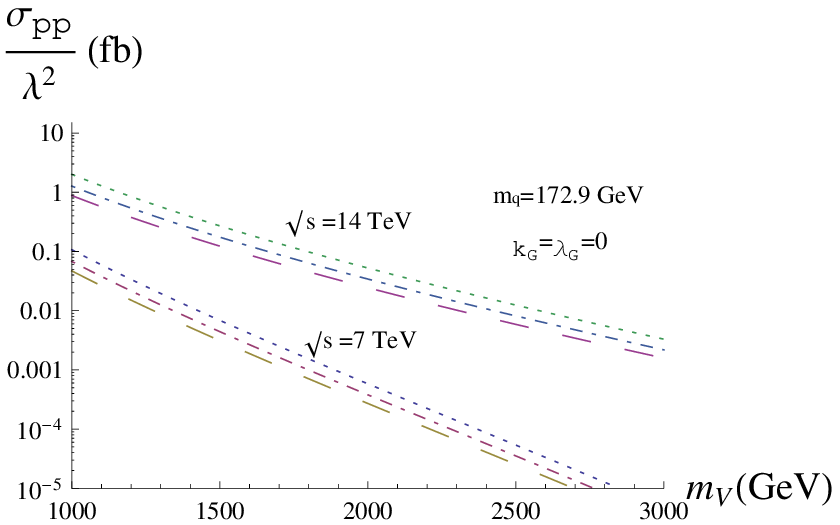}
}\caption{}
\label{fig11}
\end{figure}

Figures~\ref{fig10} and~\ref{fig11} show the cross sections for the production of the ($lu$) and ($lt$)-type vector leptoquarks ($l=e,\mu,\tau$) in $pp$~collisions evaluated using~(\ref{convolution2}) at $\sqrt{s}=7$ TeV and $\sqrt{s}=14$ TeV, $k_G=\lambda_G=0$.

\begin{figure}
\centering
\resizebox{0.9\textwidth}{!}{%
\includegraphics{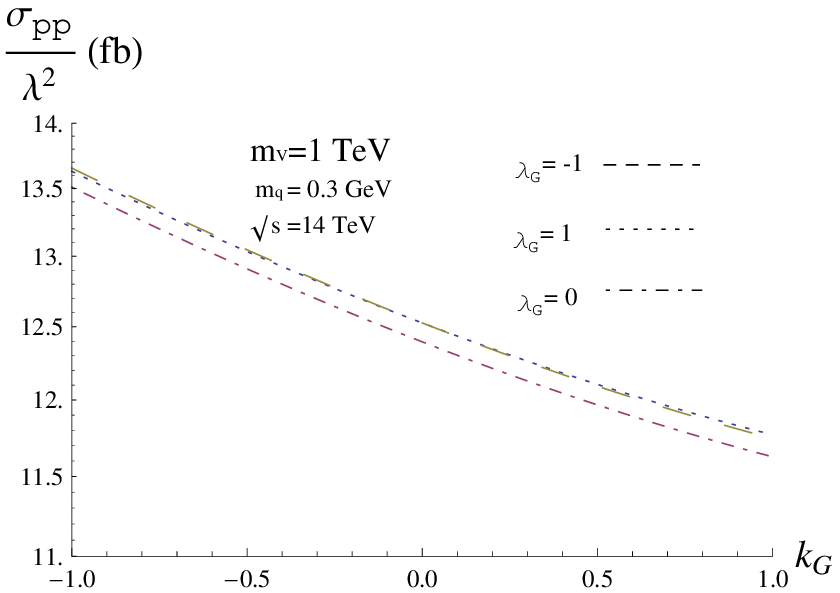}
}\caption{}
\label{fig12}
\end{figure}

\begin{figure}
\centering
\resizebox{0.9\textwidth}{!}{%
\includegraphics{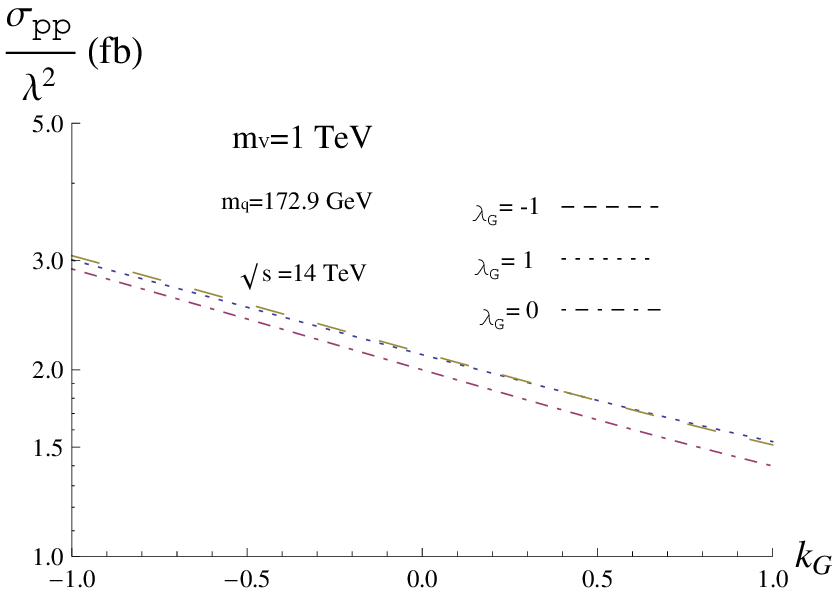}
}\caption{}
\label{fig13}
\end{figure}

\begin{figure}
\centering
\resizebox{0.9\textwidth}{!}{%
\includegraphics{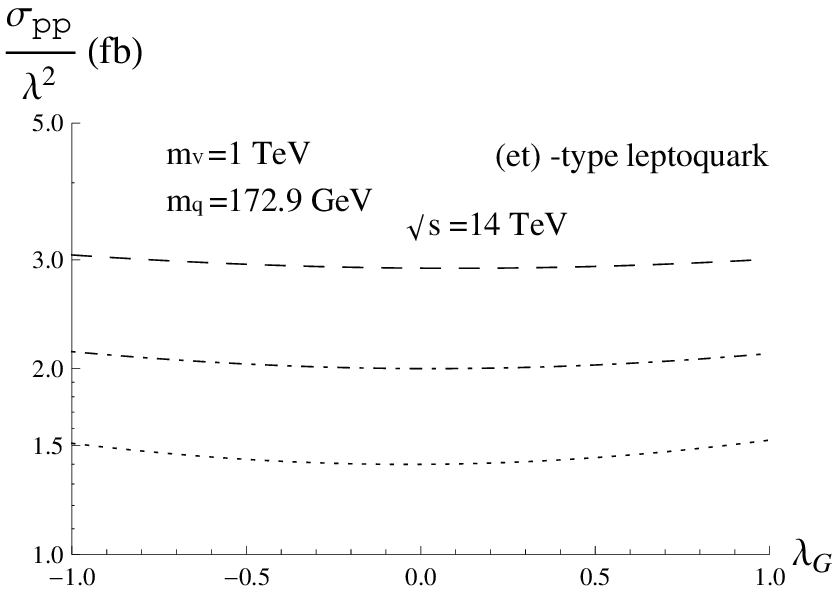}
}\caption{}
\label{fig14}
\end{figure}

Dependences of the cross sections on the anomalous couplings are also investigated in the range $|k_G|\leq1$ for $\lambda_G=1,0,-1$. The corresponding results for production of ($eu$) and ($et$)-type vector leptoquarks of mass 1~TeV in $pp$~collisions at $\sqrt{s}=14$~TeV are presented in Fig. \ref{fig12} and Figs.~\ref{fig13},~\ref{fig14} respectively.

\section{Conclusions}
The cross section for single vector leptoquark production in direct lepton--gluon interaction is calculated using the Buchm\"uller--R\"uckl--Wyler  $SU(3)_C\times SU(2)_L\times U(1)_Y$-symmetric Lagrangian with the most general $C$ and $P$ conserving coupling of the leptoquarks to gluons. An analytical expression is derived for the cross section in  the case of anomalous vector leptoquark
couplings, $k_G$ and $\lambda_G$, to the gluon field. The leptoquark mass dependences of the cross sections of inclusive production of the ($eq$), ($\mu q$) and ($\tau q$) vector leptoquarks in electron--proton collisions at $\sqrt{s}=1.8$~TeV and proton--proton collisions at the LHC energies $\sqrt{s}=7$~TeV, $\sqrt{s}=14$~TeV are evaluated for $q=u, b, t$. The dependences of the cross sections  on $k_G$ and $\lambda_G$ are also investigated in the range $|k_G|$, $|\lambda_G|\leq1$. It is found that the change of the cross sections with varying $k_G$ is sizable, while they appear to be almost insensitive to the variations of $\lambda_G$. 

The presented analysis is  applicable to the production of vector leptoquarks of the other types as well and can be useful for studies at $ep$ colliders and the LHC.

Single leptoquark production may be relatively easy to analyze experimentally due to a leptoquark decays into a lepton ($l^\pm$ or $\nu_l$) plus a quark, depending on its electric charge and isospin. In this case, the final state has simple topology  formed by either an identified charged lepton plus jet, both at large transverse momenta ($p_T\sim m_V/2$), or large missing transverse momentum plus jet~\cite{zerwas94}. 

The main contribution to the background arises from $W(\rightarrow l\nu_l)$ plus jet and $Z(\rightarrow\nu_l\bar\nu_l)$ plus jet final states. Unlike in leptoquark decays, the transverse momenta of leptons and jets are not balanced in the background events and the latter can be well separated out. 

The most stringent limits to date on the first generation vector leptoquark masses and the Yukawa coupling $\lambda$ are obtained by the ZEUS Collaboration at HERA in $e^{\pm}p$ collisions. Leptoquarks with $\lambda=0.3$ are excluded for masses up to 699 GeV~\cite{hera}. For larger values of $m_V$ ($m_V=1$ TeV) the coupling $\lambda$ may range from 0.43 to 3.24 which covers the corresponding parameter values chosen in this Letter.

The dominant mechanism of leptoquark production in $p\bar p$ and $pp$ collisions, recently analyzed at Fermilab's Tevatron and the LHC (at $\sqrt{s}=1.96$ TeV and 7 TeV, respectively) is the production of leptoquark pairs in gluon--gluon fusion and quark--antiquark annihilation~\cite{exp3,lhc1,lhc2,lhc3}. Though these processes do not depend on the unknown coupling $\lambda$, they, however, allow to reach the leptoquark masses lower than it is possible in single leptoquark production discussed above~\cite{zerwas94}. This is a result of the softness of gluons and antiquarks in the proton (antiproton). 

According to the present analysis, for the parameters given above, ($lt$)-type vector leptoquarks of mass of about 1 TeV can be reached in $pp$ collisions at $\sqrt{s}=14$ TeV with integrated luminosity of about  10 fb$^{-1}$. The ($lu$)-, ($ld$)-type vector leptoquarks of the same mass can be studied with much lower luminosities of about 1 fb$^{-1}$.

\vskip 0.5cm

{\bf Acknowledgements}
\vskip 0.5cm

This work was supported in part by the Russian Foundation for Basic Research (grant 11-02-12043), by the Program for Basic Research of the Presidium of the Russian Academy of Sciences "Fundamental Properties of Matter and Astrophysics" and by the Federal Target Program  of the Ministry of Education and Science of Russian Federation "Research and Development in Top Priority Spheres of Russian Scientific and Technological Complex for 2007-2013" (contract No. 16.518.11.7072).


\newpage

{\bf Figure Captions}
\vskip 0.5 cm 

{\bf Fig. 1:} Tree level Feynman diagrams describing the process $l+\text{g}\rightarrow q+V$.

{\bf Fig. 2:} Triple vertex $VV\text{g}$. All momenta are incoming.

\vskip 0.5 cm

{\bf Fig. 3:} The Feynman rules for the the $lqV$ vertex (top) and for the quark--gluon vertex (bottom). 

\vskip 0.5 cm

{\bf Fig. 4:} The cross section of production of the $(eb)$-type vector leptoquark in the reaction $e_Lp\rightarrow VX$ as a function of the leptoquark mass at $\sqrt{s}=1800$~GeV. Dotted curve: $k_G=\lambda_G=0$. Dotdashed curve: $k_G=1$, $\lambda_G=0$. Dashed curve: $k_G=\lambda_G=-1$.  The coupling $\lambda$ is divided out, $\alpha_s=0.118$.

\vskip 0.5 cm

{\bf Fig. 5:} The cross section of production of the $(et)$-type vector leptoquark in the reaction $e_Lp\rightarrow VX$ as a function of the leptoquark mass at $\sqrt{s}=1800$~GeV. Dotted curve: $k_G=\lambda_G=0$. Dot-dashed curve: $k_G=1$, $\lambda_G=0$. Dashed curve: $k_G=\lambda_G=-1$.  The coupling $\lambda$ is divided out, $\alpha_s=0.118$.

\vskip 0.5 cm

{\bf Fig. 6:} The cross section of production of the $(eb)$-type vector leptoquark of mass $800$ GeV in the reaction $e_Lp\rightarrow VX$ as a function of the anomalous coupling $k_G$ at $\sqrt{s}=1800$~GeV for $\lambda_G=-1$ (dashed), $\lambda_G=0$ (dot-dashed) and $\lambda_G=1$ (dotted). The coupling $\lambda$ is divided out, $\alpha_s=0.118$. 

\vskip 0.5 cm

{\bf Fig. 7:} The cross section of production of the $(et)$-type vector leptoquark of mass $800$ GeV in the reaction $e_Lp\rightarrow VX$ as a function of the anomalous coupling $k_G$ at $\sqrt{s}=1800$~GeV for $\lambda_G=-1$ (dashed), $\lambda_G=0$ (dot-dashed) and $\lambda_G=1$ (dotted). The coupling $\lambda$ is divided out, $\alpha_s=0.118$. 

\vskip 0.5 cm

{\bf Fig. 8:} The cross section of production of the $(et)$-type vector leptoquark of mass $800$ GeV in the reaction $e_Lp\rightarrow VX$ as a function of the anomalous coupling $\lambda_G$ at $\sqrt{s}=1800$~GeV for $k_G=-1$ (dashed), $k_G=0$ (dot-dashed) and $k_G=1$ (dotted). The coupling $\lambda$ is divided out, $\alpha_s=0.118$.

\vskip 0.5 cm

{\bf Fig. 9:} Schematic representation of the mechanism for producing single leptoquarks in
proton–-proton collisions.

\vskip 0.5 cm

{\bf Fig. 10:} The cross sections of production of the $(lu)$-type vector leptoquarks in the reaction $pp\rightarrow VX$ as a function of the leptoquark mass (($eu$) --dotted, $(\mu u)$ -- dot-dashed, $(\tau u)$ -- dashed). Two groups of curves are presented corresponding to $\sqrt{s}=7$ TeV (lower) and $\sqrt{s}=14$ TeV (upper). The coupling $\lambda$ is divided out, $\alpha_s=0.118$, $k_G=\lambda_G=0$.

\vskip 0.5 cm

{\bf Fig. 11:} The cross sections of production of the $(lt)$-type vector leptoquarks in the reaction $pp\rightarrow VX$ as a function of the leptoquark mass (($et$) --dotted, $(\mu t)$ -- dot-dashed, $(\tau t)$ -- dashed). Two groups of curves are presented corresponding to $\sqrt{s}=7$ TeV (lower) and $\sqrt{s}=14$ TeV (upper). The coupling $\lambda$ is divided out, $\alpha_s=0.118$, $k_G=\lambda_G=0$.

\vskip 0.5 cm

{\bf Fig. 12:} The cross section of production of the $(eu)$-type vector leptoquark of mass $1$ TeV in the reaction $pp\rightarrow VX$ as a function of the anomalous coupling $k_G$ at $\sqrt{s}=14$~TeV for $\lambda_G=-1$ (dashed), $\lambda_G=0$ (dot-dashed) and $\lambda_G=1$ (dotted). The coupling $\lambda$ is divided out, $\alpha_s=0.118$.

\vskip 0.5 cm

{\bf Fig. 13:} The cross section of production of the $(et)$-type vector leptoquark of mass $1$ TeV in the reaction $pp\rightarrow VX$ as a function of the anomalous coupling $k_G$ at $\sqrt{s}=14$~TeV for $\lambda_G=-1$ (dashed), $\lambda_G=0$ (dot-dashed) and $\lambda_G=1$ (dotted). The coupling $\lambda$ is divided out, $\alpha_s=0.118$.

\vskip 0.5 cm

{\bf Fig. 14:} The cross section of production of the $(et)$-type vector leptoquark of mass 1 TeV in the reaction $pp\rightarrow VX$ as a function of the anomalous coupling $\lambda_G$ at $\sqrt{s}=14$~TeV for $k_G=-1$ (dashed), $k_G=0$ (dot-dashed) and $k_G=1$ (dotted). The coupling $\lambda$ is divided out, $\alpha_s=0.118$.


\newpage

\end{document}